\documentclass[12pt]{article}

\usepackage{graphicx}
\usepackage{amsmath}

\input{tcilatex}

\begin{document}

\begin{titlepage}

    \title{Signs of the cusps in binary lenses}

\author{V.Bozza\thanks
{E-mail: valboz@sa.infn.it}}
\date{\empty}
\maketitle \centerline{\em $^a$ Dipartimento di Scienze Fisiche
``E.R. Caianiello'', Universit\`a di Salerno, Italy.}

 \centerline{\em $^b$ Istituto Nazionale di Fisica Nucleare, Sezione di Napoli.}

\bigskip
\begin{abstract}

The cusps of the caustics of any gravitational lens model can be
classified into positive and negative ones. This distinction lies
on the parity of the images involved in the creation/destruction
of pairs occurring when a source crosses a caustic in a cusp. In
this paper, we generalize the former definition of the sign of the
cusps. Then we apply it to the binary lens. We demonstrate that
the cusps on the axis joining the two lenses are positive while
the others are negative. To achieve our objective, we combine
catastrophe theory, usually employed in the derivation of the
properties of caustics, with perturbative methods, in order to
simplify calculations and get readable results. Extensions to
multiple lenses are also considered.

\end{abstract}

\thispagestyle{empty}  PACS: 98.62.Sb,
97.80.-d

     \vfill
     \end{titlepage}

\section*{I. Introduction}

The application of catastrophe theory in the study of the lens
mapping represents a considerable step in the understanding of the
critical behaviour in gravitational lensing \cite{Blandford &
Narayan}. Particularly interesting is the classification of
singularities through these methods, relying on the evaluation of
intrinsic mathematical quantities \cite{Kakigi Okamura &
Fukuyama}. A complete treatment of these topics can be found in
\cite{Schneider Ehlers & Falco}.

The methods of catastrophe theory have proved very powerful in the
study of the caustic structure of the binary lens \cite{Erdl &
Schneider} where it has been employed to find the transitions
between different topologies, previously studied in the equal mass
case  \cite{Schneider & Weiss a}. Erdl \& Schneider derived the
positions of the cusps and showed that, in these transitions,
pairs of cusps are created or destroyed passing through
beak--to--beak singularities \cite{Erdl & Schneider}. Witt \&
Petters \cite{Witt & Petters} used complex notation to study the
singularities of the binary lens with an additional shear field
and continuously distributed matter.

Since the binary lens is one of the most important lens models, a
detailed study of its singularities can help in the interpretation
of the physical behaviour of this system and gain information
about features that cannot be calculated analytically. For
example, the study of the amplification map near the cusps,
performed in general by Schneider \& Weiss \cite{Schneider & Weiss
b}, can provide very useful information on the amplification of
images in some regimes. Several results on cusp counting in multiple lens
systems can be found in \cite{Petters, Petters & Witt}.

A particular problem is the creation of images in the
neighbourhood of cusps. When a source crosses a caustic in a fold
singularity, two images of opposite parities are created. The
creation of these images happens in a different way when the
crossing occurs at a cusp. In this case, one pre--existing image
changes parity and two new images of the same parity are created.
As the parity of the two new images is the same of the first
before the crossing, the sum of the final parities equals the
parity of the single image before the crossing.

The parity of the original image involved in this process is a
characteristic property of the cusp, called sign \cite{Blandford &
Narayan}. Positive and negative cusps clearly behave in an
opposite way, but also the lens mapping in their neighbourhood is
influenced in different ways. The sign of a cusp can be determined
by a detailed study of the analytical form of the caustic, through
the evaluation of the fundamental quantities of catastrophe
theory.

In this work, we first give an intrinsic definition of the sign of
the cusps and then we use it to determine the signs of the cusps
in the caustics of the binary lens. Instead of dealing with the
involved exact formulae, we prefer to prove our assertions in some
particular cases where perturbative approximations are available
\cite{Bozza 2000a,Bozza 2000b} and then extend our results by
means of continuity arguments.

In Sect. II, we review the principal steps in the description of a
cusp by catastrophe theory to state our definition. Sect. III
contains the body of the calculation of the signs of the cusps in
binary lenses. Sect. IV contains some considerations about the
extensions to multiple lenses.

\section*{II. Cusps in catastrophe theory}

As usual, we introduce the Einstein radius of a reference mass
$M_0$:
\begin{equation}
R_{\mathrm{E}}^{\mathrm{0}}=\sqrt{\frac{4GM_0 }{c^{2}}
\frac{D_{\mathrm{LS}}D_{\mathrm{OL}}}{D_{\mathrm{OS}}}}.
\end{equation}
We indicate the coordinates in the lens plane normalized to
$R_\mathrm{E}^0$ by $\mathbf{x}=\left( x_1; x_2 \right)$ and the
coordinates in the source plane by $\mathbf{y}= \left( y_1; y_2
\right)$. All masses are measured in terms of $M_0$. The matter
density normalized to the critical density
\begin{equation}
\Sigma_\mathrm{cr}=\frac{c^2D_\mathrm{OS}}{4\pi G D_{OL}
D_\mathrm{LS}}
\end{equation}
is $\kappa \left( \mathbf{x} \right) $.

The Fermat potential of a given distribution of matter is:
\begin{equation}
\phi \left( \mathbf{x}, \mathbf{y} \right)=\frac{1}{2} \left(
\mathbf{x}-\mathbf{y} \right)^2-\frac{1}{\pi} \int \mathrm{d}^2 x'
\kappa \left( \mathbf{x}' \right) \ln \left|
\mathbf{x}-\mathbf{x}' \right|.
\end{equation}
The lens equation is obtained by taking the gradient of this
potential:
\begin{equation}
\nabla_\mathbf{x} \phi \left( \mathbf{x}, \mathbf{y} \right)=0.
\end{equation}
This equation can be written in the form of an application from
the lens plane to the source plane:
\begin{equation}
\mathbf{y}=\mathbf{y} \left( \mathbf{x} \right). %
\label{Lens mapping}
\end{equation}

Given a source position $\mathbf{y}$, the $\mathbf{x}$'s solving
the equation (\ref{Lens mapping}) are called images.

The local properties of the lens mapping can be studied through
its Jacobian matrix
\begin{equation}
J=\left( \begin{array}{cc}
  \frac{\partial y_1}{\partial x_1} & \frac{\partial y_1}{\partial x_2} \\
  \frac{\partial y_2}{\partial x_1} & \frac{\partial y_2}{\partial x_2}
\end{array}\right)
=\left( \begin{array}{cc}
   1-\kappa- \gamma_1 & -\gamma_2 \\
   -\gamma_2 & 1- \kappa + \gamma_1
  \end{array}\right),
\end{equation}
where $\gamma_1=\frac{1}{2} \left( \phi_{11}- \phi_{22} \right)$
and $\gamma_2= \phi_{12}= \phi_{21} $. We use the notation
$\phi_i=\frac{\partial \phi}{\partial x_i} $. The trace of the
Jacobian matrix is
\begin{equation}
\mathrm{Tr}J=2 \left(1- \kappa \right).
\end{equation}

Many properties of the lens mapping can be studied through the
determinant $D$ of the Jacobian matrix
\begin{equation}
D=\left( 1- \kappa \right)^2- \left( \gamma_1^2+\gamma_2^2
\right).
\end{equation}
The sign of this determinant, evaluated at the position of an
image, is called parity. Images with negative parity are
characterized by a reversed handedness. In the points where $D=0$,
the lens mapping is not invertible. These points are arranged in
smooth curves called critical curves and their images through the
lens mapping (\ref{Lens mapping}) are called caustics \cite{Levine
Petters & Wambsganss}.

If a parameterization of the critical curve is available in the
form:
\begin{equation}
\mathbf{x}=\mathbf{x}^\mathrm{c} \left( \theta \right),
\end{equation}
its tangent vector can be written as:
\begin{equation}
\mathbf{T}=\frac{\mathrm{d} \mathbf{x}^\mathrm{c}}{\mathrm{d} \theta}. %
\label{Tangent vector}
\end{equation}

For the caustic, we have the natural parameterization:
\begin{equation}
\mathbf{y}=\mathbf{y}^\mathrm{c} \left( \theta
\right)=\mathbf{y}\left( \mathbf{x}^\mathrm{c} \left( \theta
\right) \right),
\end{equation}
leading to a tangent vector:
\begin{equation}
\frac{\mathrm{d} \mathbf{y}^\mathrm{c}}{\mathrm{d} \theta}=J
\mathbf{T}= \left( \mathbf{T} \cdot \nabla \right) \mathbf{y}
\left( \mathbf{x} \right)|_{\mathbf{x}=\mathbf{x}^\mathrm{c}\left(
\theta \right)}.
\end{equation}

The caustic singularities can be classified into different types,
including folds, cuspoids, umbilics, beak--to--beaks and lips
\cite{Schneider Ehlers & Falco}.

The zero order singularity is the fold, that is a simple critical
point, satisfying:
\begin{equation}
  D=0,
  \label{Fold condition}
\end{equation}
not belonging to other special classes.

What we are interested into is the cuspoid sequence. A cusp point
is a point  belonging to the critical curve (thus satisfying
condition (\ref{Fold condition})), such that the vector tangent to
the caustic vanishes:

\begin{equation}
  \frac{\mathrm{d} \mathbf{y}^\mathrm{c}}{\mathrm{d}\theta} = 0.
\label{Cusp condition}
\end{equation}
Its image on the caustic is the cusp.

The cuspoid of second order is the swallowtail, coming up when the
condition:
\begin{equation}
  \frac{\mathrm{d}^2 \mathbf{y}^\mathrm{c}}{\mathrm{d}\theta^2} = 0
  \label{Swallowtail}
\end{equation}
is added to the conditions (\ref{Fold condition}) and (\ref{Cusp
condition}). The vanishing of higher order derivatives leads to
singularities of higher order. These definitions do not depend on
reparameterizations of the critical curve, since for a cuspoid of
order $p$ we require all derivatives up to the $p^\mathrm{th}$
order to vanish simultaneously.

To analyse in detail the lensing behaviour near a cusp, it is
convenient to choose the coordinates so as to have the cusp point
in the origin of the lens plane and the cusp in the origin of the
source plane. With an opportune rotation, the lens mapping can be
generally written in the form
\begin{mathletters}
\begin{eqnarray}
&& y_1=cx_1+\frac{1}{2}bx_2^2 \\%
&& y_2=bx_1x_2+ax_2^3,
\end{eqnarray}
\end{mathletters}
where $a$, $b$, $c$ are coefficients depending on the derivatives
of the Fermat potential evaluated at the cusp point
\cite{Schneider Ehlers & Falco}.

The determinant and the trace of the Jacobian, in this frame, are
\begin{eqnarray}
&& D= bcx_1+\left(3ac-b^2 \right)x_2^2 \\%
&& \mathrm{Tr}J=c.
\end{eqnarray}

The critical curve is found solving the equation $D=0$. We
parameterize it in this way:
\begin{mathletters}
\begin{eqnarray}
& & x^c_1=\frac{b^2-3ac}{bc}\theta^2\\ %
& & x^c_2= \theta.
\end{eqnarray}
\end{mathletters}

The caustic is
\begin{mathletters}
\begin{eqnarray}
& & y^c_1=3\frac{b^2-2ac}{2b}\theta^2\\ %
& & y^c_2= \frac{b^2-2ac}{c}\theta^3.
\end{eqnarray}
\end{mathletters}

\begin{figure}
 \resizebox{\hsize}{!}{\includegraphics{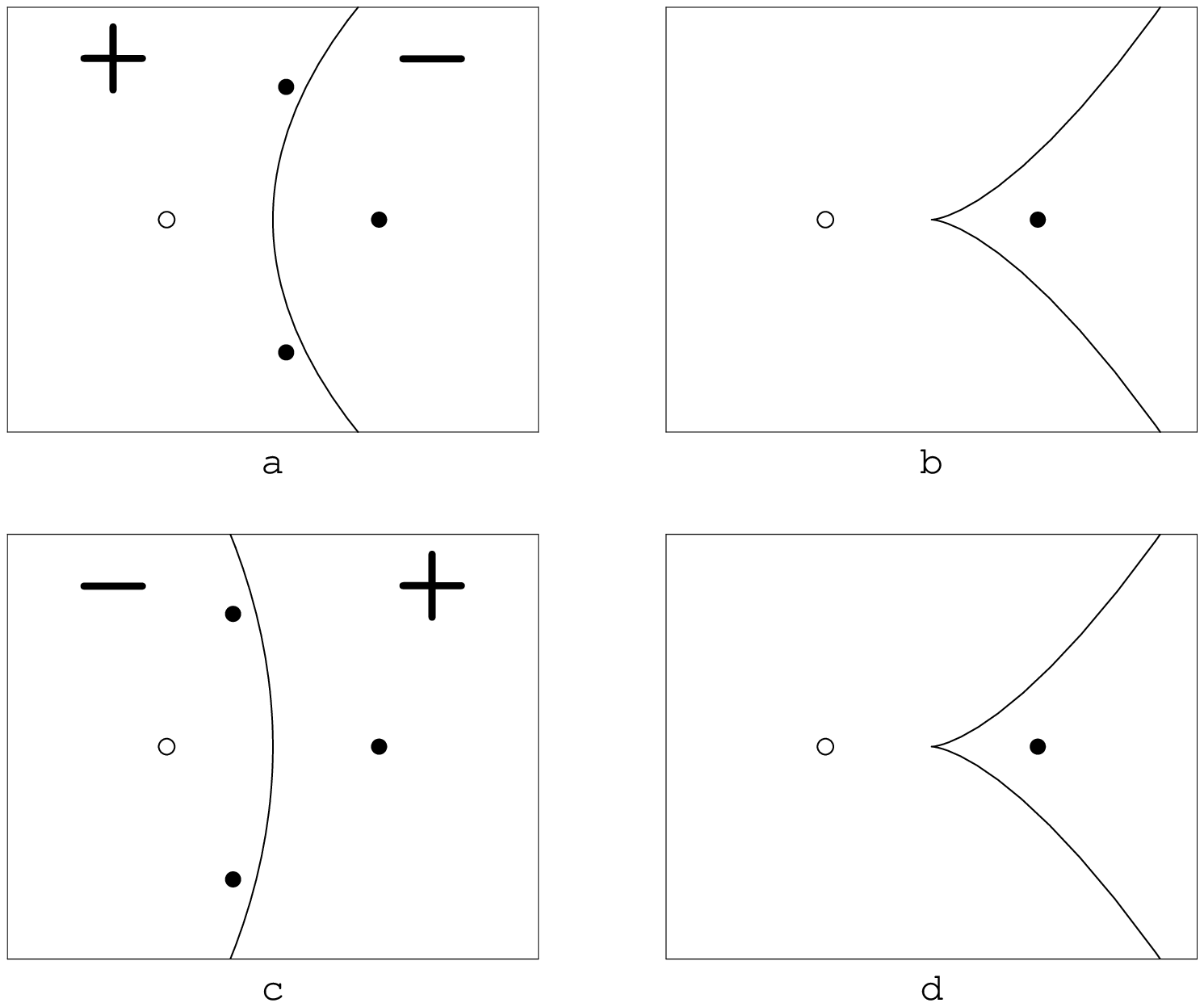}}
 \caption{The lens mapping near cusps. On the right column, %
 we have the caustics corresponding to the critical %
 curves on the left. The first row represents the situation %
 for a positive cusp. The bottom row shows the situation for %
 a negative cusp. The sign of $D$ is indicated in the left %
 figures. The circles on the left figures are the images %
 corresponding to the source positions in the right. %
 When the source is outside of the caustic (empty circles) there is %
 one image, while when it is inside (filled circles) three images %
 are formed. Their parity is discussed in the text.}
 \label{Fig definition of the sign}
\end{figure}

The sign of the quantity $2ac-b^2$ defines the sign of the cusp
\cite{Blandford & Narayan}. We can guess that the behaviour of the
caustic changes in an evident way according to this sign. To see
these differences, one can solve the lens mapping explicitly in
the neighbouhood of a cusp point. A detailed discussion of this
topic can be found in Refs. \cite{Blandford & Narayan, Schneider
Ehlers & Falco}. In Fig. \ref{Fig definition of the sign}, we
briefly recall the behaviour of the lens mapping near positive
(top row) and negative cusps (bottom row). When a source crosses
the caustic in a positive cusp, a positive image crosses a
critical curve, thus changing its parity. Moreover, two positive
images come up from the crossing point in the direction tangential
to the critical curve. For negative cusps, all parities are
reversed. Then a negative image changes its parity and a pair of
negative images arises. Of course, the sum of the parities is
unchanged in both processes.

Now we want to give a definition for the sign of the cusp that can
be used in all frames. With this definition, we will be able to
compute easily the signs of the cusps for the caustics of any lens
mapping without having to look for the standard frame for each
cusp.

The vector tangent to the cusp is
\begin{equation}
  \frac{\mathrm{d} \mathbf{y}^\mathrm{c}}{\mathrm{d}\theta} =\left(
  \begin{array}{c}
3\frac{b^2-2ac}{b}\theta \\ %
3\frac{b^2-2ac}{c}\theta^2
\end{array} \right).
\end{equation}
Its derivative, evaluated at the cusp is
\begin{equation}
\frac{\mathrm{d}^2 \mathbf{y}^\mathrm{c}}{\mathrm{d}\theta^2} =%
\left( \begin{array}{c}
3\frac{b^2-2ac}{b}\\ %
0
\end{array} \right).
\label{2nd derivative}
\end{equation}
We can notice, recalling (\ref{Swallowtail}), that a cusp with a
vanishing sign is a swallowtail.

The vector
\begin{equation}
\mathbf{N}=  -\frac{\nabla D}{\mathrm{Tr} J} %
\label{Normal vector}
\end{equation}
is everywhere orthogonal to the critical curve. In the standard
frame, it reads
\begin{equation}
\mathbf{N}=\left(\begin{array}{c}%
-b \\%
\frac{2\left(b^2-3ac\right)}{c}x_2
\end{array}\right) .
\end{equation}
At the cusp point, $\mathbf{N}$ is parallel to the second order
derivative of the caustic (\ref{2nd derivative}).

Then, the quantity
\begin{equation}
 \mathbf{N}\cdot\frac{\mathrm{d}^2 \mathbf{y}^\mathrm{c}}{\mathrm{d}\theta^2},
  \label{Sign of the cusp}
\end{equation}
evaluated at the cusp point, is $3\left(2ac-b^2 \right)$. So its
sign coincides with the sign of the cusp. Moreover, the expression
(\ref{Sign of the cusp}) is invariant for translations and
rotations of the coordinate frame and for reparameterizations of
the critical curve (as long as it is evaluated at the cusp point).
Then it is natural to take Eq. (\ref{Sign of the cusp}) as the
general definition for the sign of the cusp, valid not only in the
standard frame but for any choice of the coordinates and
parameterizations of the critical curve.

In what follows, we will use this definition to determine the
signs of the cusps in the binary lens.

\section*{III. Signs of the cusps in binary lenses}

\begin{figure}
 \resizebox{6cm}{!}{\includegraphics{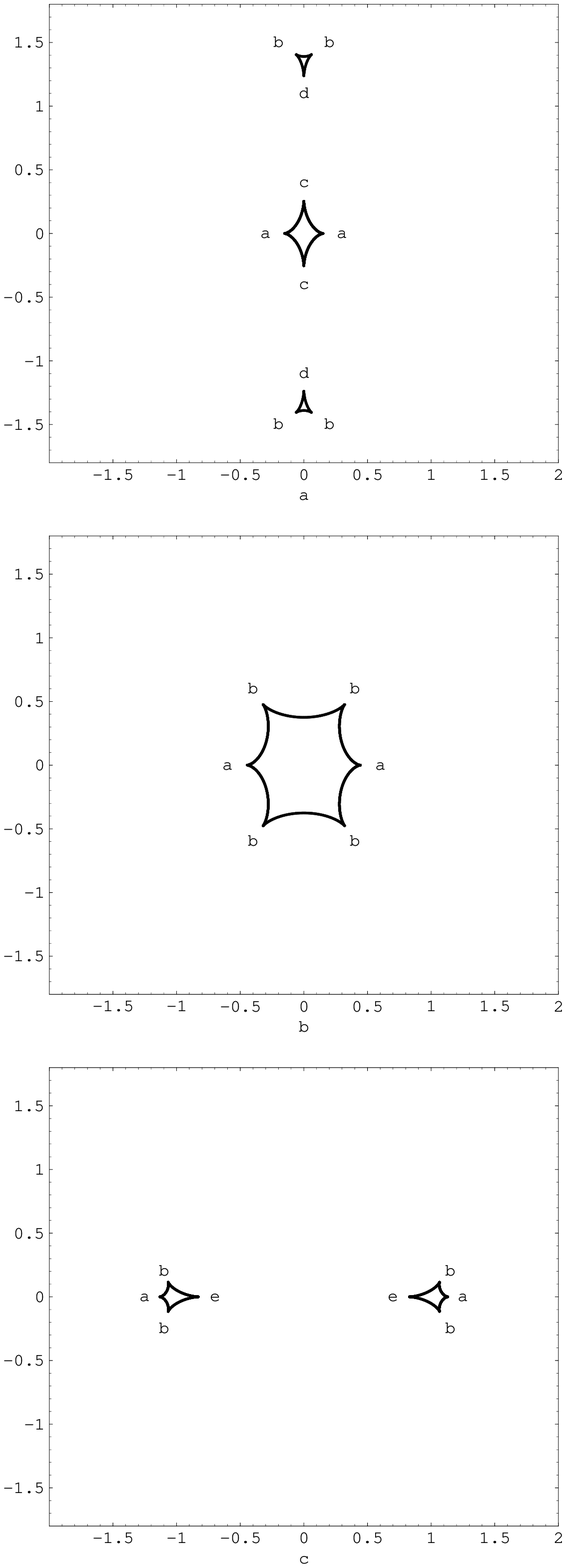}}
 \hfill
 \caption{Topologies of the caustics of the equal--mass binary lens. %
 (a) Close binary, (b) intermediate binary, (c) wide binary. %
 The letters identify the different kinds of cusps.}
 \label{Fig topologies}
\end{figure}

The critical curves of binary lenses can be classified according
to two parameters: the mass ratio of the two lenses and their
separation. We indicate the two masses (always measured in terms
of the reference mass $M_0$) by $m_1$ and $m_2$. The total mass is
$M=m_1+m_2$. For each mass ratio, three possible topologies are
present depending on the separation $a$ between the lenses. In
Fig. \ref{Fig topologies}, we show them for a system of two equal
masses.

For close binaries, i.e. when $ \frac{\left(  M ^2-a^4
\right)^3}{27 a^8}>m_1 m_2$ \cite{Witt & Petters}, three caustics
are present: the one at the centre of the system has four cusps,
while the other two have three cusps (Fig. \ref{Fig topologies}a).

Wide binary systems, characterized by the condition $a^2>\left(
m_1^{1/3}+m_2^{1/3} \right)^3$, have two caustics with four cusps
each (Fig. \ref{Fig topologies}c).

Intermediate binaries have one caustic with six cusps (Fig.
\ref{Fig topologies}b).

If we change the parameters of the system continuously, we can
follow each cusp unambiguously, since their positions are
continuous functions of the parameters. Also the quantity
(\ref{Sign of the cusp}), giving the sign of the cusp, is a
continuous function of the parameters. So, remembering that the
two vectors in the scalar product of Eq. (\ref{Sign of the cusp})
are parallel at the cusp, a change of sign in a cusp can only
occur through a higher order singularity, according to Eq.
(\ref{Swallowtail}). But no higher order cuspoid is present in
binary lenses, so each cusp bears its sign unaltered as the
parameters change, even when transitions between different
topologies occur (if the cusp is not directly involved in these
transitions).

These observations allow us to introduce an overall labelling for
the cusps as we have done in Fig. \ref{Fig topologies}. As we are
interested in the study of the sign of the cusps, we have given
the same label to the cusps which can be obtained by reflection on
one or both axes in the equal--masses case, as they obviously have
the same sign. For this reason we have two cusps of type $a$, four
of type $b$ and so on.

In total, twelve cusps can be distinguished in the binary lens:
two of them (indicated by $e$ in Fig. \ref{Fig topologies}c) are
only present in wide binaries; four of them (indicated by $c$ and
$d$ in Fig. \ref{Fig topologies}) only exist in close binaries.
The other six are always present. The disappearance of cusps $c$
and $d$ and of cusps $e$ occurs through beak--to--beak
singularities at the critical values of the separation between the
masses, leading to the intermediate topology \cite{Erdl &
Schneider}.

Since the sign of the cusp is a characteristic not depending on
the particular choice of the parameters, we can lead our study in
the simplest cases and then extend our results in general,
exploiting the continuity of the quantities involved in our
calculations. Of course, a natural simplifying choice can be the
equal--mass case, as concerns the mass ratio. So, from now on we
put $m_1=m_2=M/2$.

Now, we could deal with the exact formulae for the caustics, given
in Refs. \cite{Schneider & Weiss a, Erdl & Schneider, Witt &
Petters}, but we would rather simplify our treatment further,
reducing to the cases where perturbative results are employable.
In this way, by very few steps, we can derive properties of the
general binary lens from the study of its particular cases.

In Ref. \cite{Dominik} approximations are given for the caustics
of the binary lens in the form of multipole expansions in its
extreme cases: very close binary systems ($a \ll \sqrt{M}$), very
wide binary systems ($a \gg \sqrt{M}$) and planetary systems ($m_2
\ll m_1$). Perturbative expansions to the first significant orders
for multiple lenses were derived by Bozza \cite{Bozza 2000a} in
the same cases. In the following subsections, we use the
expressions in refs. \cite{Bozza 2000a, Bozza 2000b} to analyze
the signs of the cusps in representative cases. Deriving these
signs for all types of cusps identified in Fig. \ref{Fig
topologies}, our demonstration will be concluded.

\subsection*{A. Cusps in wide binary caustics}

Let's begin from the caustics formed by the two lenses when they
are very far each other. Choosing the mass of each single lens as
the reference mass, i.e. $M/2=1$, the starting hypothesis is
\begin{equation}
a \gg 1 . \label{Wide hypothesis}
\end{equation}

These caustics are the results of the deformations on the two
separated Einstein rings of each mass induced by the other. We put
the first lens at the origin and the second at the position
$\mathbf{x}_2=\left(a;0 \right)$. Thanks to Eq. (\ref{Wide
hypothesis}), we shall expand all our objects in series of the
perturbative parameter $\frac{1}{a}$ to the first non trivial
order. The lens equation becomes:
\begin{mathletters}
\begin{eqnarray}
y_1&=&x_1-\frac{x_1}{\left|\mathbf{x}
\right|^2}+\frac{1}{a}+\frac{x_1}{a^2}, \\ %
y_2&=&x_2-\frac{x_2}{\left|\mathbf{x} \right|^2} -\frac{
x_2}{a^2}.
\end{eqnarray}
\end{mathletters}

The determinant of the Jacobian matrix is
\begin{equation}
D=1-\frac{1}{\left|\mathbf{x} \right|^4}-\frac{2 \left(
x_1^2-x_2^2 \right)}{a^2 \left|\mathbf{x} \right|^4}. %
\label{Wide D}
\end{equation}

The solution of the equation $D=0$ with $D$ given by Eq.
(\ref{Wide D}) is the critical curve of the first mass \cite{Bozza
2000a}
\begin{equation}
\mathbf{x}^\mathrm{c} \left( \theta \right)=\left( \begin{array}{c}
  \cos \theta \left( 1+ \frac{ \cos 2\theta}{2a^2}\right) \\
  \\
  \sin \theta \left( 1+ \frac{ \cos 2\theta}{2a^2}\right)
\end{array}\right).
\end{equation}

The caustic is
\begin{equation}
\mathbf{y}^\mathrm{c} \left( \theta \right)=\left(
\begin{array}{c}
  \frac{1}{a}+\frac{2\cos^3 \theta}{a^2}  \\
  \\
  -\frac{2\sin^3 \theta}{a^2}
\end{array}\right) .
\end{equation}

Now we can build the objects involved in the classification of the
cuspoids. The vector tangent to the caustic
\begin{equation}
    \frac{\mathrm{d} \mathbf{y}^\mathrm{c}}{\mathrm{d}\theta} =
    -\frac{3 \sin 2\theta}{a^2}\left(
    \begin{array}{c}
   \cos \theta \\
   \\
   \sin \theta
\end{array}\right)
\end{equation}
vanishes at $\theta=n\pi/2$ for $n=0,1,2,3$. The cusp obtained for
$n=0$ is on the $x_1$--axis, directed towards the other mass.
Looking at Fig. \ref{Fig topologies}c, where the caustic of the
first mass is the left one, it can be identified as a cusp of type
$e$. The cusp for $n=2$ is in the opposite direction and is a cusp
$a$. The other two are cusps of type $b$.

Now, let's calculate the signs of these cusps. The second
derivative of the caustic is
\begin{equation}
    \frac{\mathrm{d}^2 \mathbf{y}^\mathrm{c}}{\mathrm{d}\theta^2}  =-\frac{3 }{a^2}\left( \begin{array}{c}
   \cos \theta \left( -1+ 3 \cos 2\theta \right) \\
   \\
   \sin \theta \left( 1+ 3 \cos 2\theta \right)
\end{array}\right),
\end{equation}
and the vector $\mathbf{N}$, defined in Eq. (\ref{Normal vector})
is:
\begin{equation}
\mathbf{N}=\left( \begin{array}{c}
   -2\cos \theta+\frac{1}{a^2} %
   \left( 5 \cos \theta + \cos 3\theta \right) \\
   \\
   -2\sin \theta+\frac{1}{a^2} %
   \left( - 5 \cos \theta + \cos 3\theta \right)
\end{array} \right).
\end{equation}

Finally, the quantity (\ref{Sign of the cusp}) evaluated at the
four cusps is
\begin{equation}
\left. \mathbf{N}\cdot \frac{\mathrm{d}^2
\mathbf{y}^\mathrm{c}}{\mathrm{d}\theta^2}
\right|_{\theta=n\pi/2}=\left( -1 \right)^n \frac{12 }{a^2}.
\end{equation}

Then, the cusps $a$ and $e$ are positive, while the cusps $b$ are
negative. Having established the signs of these three types of
cusps, the only cusps to be studied are of type $c$ and $d$, to be
found in the analysis of close binary systems.

\subsection*{B. Cusps in close binary main caustic}

We calculate now the signs of the cusps of the main caustic of
close binary systems. This is necessary to obtain the sign of the
cusps of type $c$. However, we also get a confirmation for cusps
of type $a$. Now we choose the origin to be at the center between
the two masses and consider the total mass as the reference mass:
$M=1$, $\mathbf{x}_1=\left( -\frac{a}{2};0 \right)$,
$\mathbf{x}_2=\left( \frac{a}{2};0 \right)$. We consider the limit
\begin{equation}
a \ll 1.
\end{equation}

By virtue of this hypothesis, all objects will be expanded in
powers of $a$. Stopping at the first significant order, the lens
equation becomes
\begin{mathletters}
\begin{eqnarray}
y_1&=&x_1-\frac{ x_1}{\left|\mathbf{x} \right|^2}-\frac{ a^2 x_1
\left( x_1^2-3 x_2^2 \right)}{4\left|\mathbf{x} \right|^6} \\ %
y_2&=&x_2-\frac{x_2}{\left|\mathbf{x} \right|^2} +\frac{a^2 x_2
\left( x_2^2-3 x_1^2\right)}{4\left|\mathbf{x} \right|^6}.
\end{eqnarray}
\end{mathletters}

The Jacobian determinant is
\begin{equation}
D=1-\frac{1}{\left|\mathbf{x} \right|^4}-\frac{3 a^2\left(
x_1^2-x_2^2 \right)}{2 \left|\mathbf{x} \right|^8}. %
\label{Close D}
\end{equation}

The main critical curve is given by the expression \cite{Bozza
2000a}
\begin{equation}
\mathbf{x}^\mathrm{c} \left( \theta \right)=\left(
\begin{array}{c}
   \cos \theta \left( 1+ \frac{3 a^2 \cos 2\theta}{8}\right) \\
  \\
   \sin \theta \left( 1+ \frac{3 a^2\cos 2\theta}{8}\right)
\end{array}\right).
\end{equation}

The caustic is
\begin{equation}
\mathbf{y}^\mathrm{c} \left( \theta \right)=\left(
\begin{array}{c}
  \frac{a^2}{2}\cos^3 \theta  \\
  \\
  -\frac{a^2}{2}\sin^3 \theta
\end{array}\right) .
\end{equation}

The calculation proceeds in a way analogous to the previous case.

The vector tangent to the caustic
\begin{equation}
  \frac{\mathrm{d}
\mathbf{y}^\mathrm{c}}{\mathrm{d}\theta} =-\frac{3 a^2 \sin
2\theta}{4}\left( \begin{array}{c}
   \cos \theta \\
   \\
   \sin \theta
\end{array}\right)
\end{equation}
vanishes at $\theta=n\pi/2$ for $n=0,1,2,3$. The cusps obtained
for $n=0,2$ are on the $x_1$--axis, thus being of type $a$. The
other two are cusps of type $c$.

The second derivative is
\begin{equation}
  \frac{\mathrm{d}^2
\mathbf{y}^\mathrm{c}}{\mathrm{d}\theta^2} =-\frac{3 a^2}{8}\left(
\begin{array}{c}
   \cos \theta + 3 \cos 3\theta \\
   \\
   -\sin \theta + 3 \sin 3\theta
\end{array}\right)
\end{equation}
and the vector $\mathbf{N}$ is
\begin{equation}
\mathbf{N}=\left( \begin{array}{c}
   -2\cos \theta+\frac{3a^2}{8} %
   \left( \cos \theta -3 \cos 3\theta \right) \\
   \\
   -2\sin \theta-\frac{3a^2}{8} %
   \left( \sin \theta +3 \sin 3\theta \right)
\end{array} \right).
\end{equation}

Finally, the quantity (\ref{Sign of the cusp}) evaluated at the
four cusps is:
\begin{equation}
\left. \mathbf{N}\cdot \frac{\mathrm{d}^2
\mathbf{y}^\mathrm{c}}{\mathrm{d}\theta^2}
\right|_{\theta=n\pi/2}=\left( -1 \right)^n \frac{3 a^2}{2}.
\end{equation}

Then, besides the confirmation of the positivity of cusps $a$, we
find that the cusps $c$ are negative.

\subsection*{C. Cusps in close binary secondary caustics}

Finally, we need the sign of the cusps of type $d$. They arise in
close binary systems and belong to secondary caustics. Studying
the cusps of these caustics, we also have the confirmation of the
negativity of cusps $b$. We retain the same choices of the
previous subsection for the parameters of the system. The critical
curve we investigate is the upper secondary one \cite{Bozza 2000b}
\begin{equation}
\mathbf{x}^\mathrm{c} \left( \theta \right)=\left(
\begin{array}{c}
   \frac{a^3 \cos \theta}{4} \\
   \\
   \frac{a}{2}+\frac{a^3 \sin \theta}{4}
\end{array}\right).
\end{equation}
corresponding to the bottom secondary caustic
\begin{equation}
\mathbf{y}^\mathrm{c} \left( \theta \right)=\left(
\begin{array}{c}
  \frac{a^3 }{8} \left(2\cos \theta- \sin 2\theta  \right)  \\
  \\
  -\frac{1}{a}+\frac{a}{2}+\frac{a^3 }{8} \left(2\sin \theta- \cos
  2\theta\ \right)
\end{array}\right) .
\end{equation}

The vector tangent to the caustic
\begin{equation}
\frac{\mathrm{d} \mathbf{y}^\mathrm{c}}{\mathrm{d}\theta} =\frac{
a^3 }{4}\left(
\begin{array}{c}
   \cos 2\theta+ \sin \theta \\
   \\
   \sin 2\theta+ \cos \theta
\end{array}\right)
\end{equation}
vanishes at $\theta=2n\pi/3+ \pi/2$ for $n=0,1,2$. The cusp
obtained for $n=0$ is on the $x_2$--axis, then it is the one of
type $d$. The other two are the cusps of type $b$.

The second order derivative is:
\begin{equation}
 \frac{\mathrm{d}^2
\mathbf{y}^\mathrm{c}}{\mathrm{d}\theta^2} =\frac{a^3}{4}\left(
\begin{array}{c}
   \cos \theta - 2 \sin 2\theta \\
   \\
   -\sin \theta + 2 \cos 2\theta
\end{array}\right)
\end{equation}
and the vector $\mathbf{N}$ is
\begin{equation}
\mathbf{N}=\left( \begin{array}{c}
   \frac{4\cos \theta}{a^3} \\
   \\
   \frac{4\sin \theta}{a^3}\end{array} \right).
\end{equation}

Finally, the quantity (\ref{Sign of the cusp}) evaluated at the
four cusps is:
\begin{equation}
\left. \mathbf{N}\cdot \frac{\mathrm{d}^2
\mathbf{y}^\mathrm{c}}{\mathrm{d}\theta^2}
\right|_{\theta=2n\pi/3+\pi/2}=-3 \; \; \; \forall n.
\end{equation}

All of the cusps are negative, not only the ones of type $b$,
already studied, but also the cusp $d$.

In conclusion, in binary lenses, we have that the cusps on
$x_1$--axis ($a$ and $e$) are positive and the others are
negative.

\section*{IV. Extensions to multiple systems}

In refs. \cite{Bozza 2000a, Bozza 2000b}, the expressions for the
caustics were derived not only for binary systems but for a
general multiple lens. Then, by further continuity arguments, we
can extend our results even to some special multiple lenses.

In a first approximation \cite{Bozza 2000a}, the caustic of a lens
very far from the other components of the distribution of mass is
a Chang--Refsdal caustic  \cite{Chang & Refsdal 1, Chang & Refsdal
2}, with shear:
\begin{equation}
\gamma=\sqrt{\left[ \sum\limits_{i=2}^n \frac{m_i \sin \left(
2\varphi_i \right)}{\rho_i^2}\right]^2+\left[ \sum\limits_{i=2}^n
\frac{m_i \cos \left( 2\varphi_i \right)}{\rho_i^2}\right]^2},%
\label{Wide caustic shear}
\end{equation}
along the direction
\begin{equation}
\varphi=\frac{1}{2}\arctan \left[ \frac{\sum\limits_{i=2}^n
\frac{m_i \sin \left( 2\varphi_i
\right)}{\rho_i^2}}{\sum\limits_{i=2}^n \frac{m_i \cos \left(
2\varphi_i \right)}{\rho_i^2}} \right].%
\label{Wide caustic angle}
\end{equation}
In these two expressions, the positions of the masses producing
the shear are given in polar coordinates $\rho_i$, $\varphi_i$.

In Sect. 3.1, we have analyzed the case of a binary system where
only the mass $m_2$ produced the shear field on the first mass.
This field was directed towards $m_2$. We found that the cusps
along this direction were positive and the other two were
negative. Then, if we let a third mass $m_3$ arise from zero to
some finite value (also placed far from the mass $m_1$, in order
to remain in the approximation of wide multiple lenses), the shear
would be modified, but the shape of the caustic would remain the
same. The positions of the cusps change continuously and then we
can still say that the cusps on the shear axis are positive and
the other two are negative as long as the topology is unaltered.

Regarding close multiple lenses, we can follow the same procedure.
In this case, the caustic formed by masses having mutual distances
much smaller than the total Einstein radius is a quadrupole
caustic \cite{Bozza 2000a}, with moment
\begin{equation}
Q=M \sqrt{\left[ \sum\limits_{i=1}^n m_i \rho_i^2 \cos \left( 2
\varphi_i \right) \right]^2+\left[ \sum\limits_{i=1}^n m_i
\rho_i^2 \sin \left( 2 \varphi_i \right) \right]^2 },
\end{equation}
oriented along
\begin{equation}
\varphi=\arctan \frac{Q-M\sum\limits_{i=1}^n m_i \rho_i^2 \cos
\left( 2 \varphi_i \right)}{M\sum\limits_{i=1}^n m_i \rho_i^2 \sin
\left( 2 \varphi_i \right)}.
\end{equation}

In the binary case, examined in Sect. 3.2, $\varphi=0$. The two
cusps on this axis were positive and the other two were negative.
Again, turning on additional masses from zero to some finite
value, the orientation changes continuously and the signs of the
cusps are preserved as long as the topology remains the same.
Then, the cusps on the orientation axis are positive and the other
two are negative.

For secondary caustics, in close multiple systems, we start
referring to planetary systems. If we have a star surrounded by
one planet placed inside the Einstein ring of the star, two
secondary critical curves are formed very close to the planet
\cite{Bozza 1999}. According to the arguments in Sect. 3.3, the
resulting secondary caustics have three negative cusps each. Now,
if we add more planets inside the Einstein ring of the star, each
of them will be accompanied by two small secondary critical
curves. Since, in a first approximation, each couple of secondary
critical curves formed by one planet is not affected by the
presence of the other planets \cite{Bozza 1999}, we deduce that
these caustics have the same properties of the binary secondary
caustics (obtained putting the masses of the other planets to
zero): they have only negative cusps. Now, we let the masses of
all the planets increase continuously from their small values to
their definitive values, in order to obtain a generic multiple
system with distances smaller than the global Einstein radius. The
positions of the secondary critical curves will change, but their
shape (and the signs of the cusps) will remain the same as long as
they remain simple critical curves, i.e. as long as they do not
meet each other to give multiple critical curves \cite{Bozza
2000b}. Yet, even in this case, the transitions can be achieved by
suppressing two cusps through a beak--to--beak singularity
\cite{Bozza 2000b}. The surviving cusps will anyway preserve their
sign. In conclusion, all secondary caustics (both simple and
multiple) have negative cusps. A direct calculation using the
formulae given in \cite{Bozza 2000b} confirms this statement
derived here by continuity arguments.

We see that the approach used in this specific problem
(calculations in perturbative cases and extensions by continuity
arguments) can be a powerful tool in order to gain information
about situations where there is poor analytical direct knowledge.
In the study of caustics of gravitational lens models, this
methodology can surely provide a great help for the comprehension
of the topologies actually realized in physical systems.

\bigskip
\bigskip

\begin{centerline}
{\bf Acknowledgments}
\end{centerline}

I would like to thank Salvatore Capozziello for his helpful
comments on the manuscript.

Work supported by fund ex 60\% D.P.R. 382/80.

\newpage

\newpage

\end{document}